\documentclass[twocolumn,english,aps,prl]{revtex4}
\usepackage{lmodern}
\usepackage{lmodern}
\usepackage[T1]{fontenc}
\usepackage[latin9]{inputenc}
\usepackage{color}
\usepackage{babel}
\usepackage{mathrsfs}
\usepackage{amsmath}
\usepackage{amssymb}
\usepackage{graphicx}
\usepackage{esint}
\usepackage[unicode=true,pdfusetitle,
 bookmarks=false,
 breaklinks=false,pdfborder={0 0 1},backref=false,colorlinks=true]
 {hyperref}
\hypersetup{
 colorlinks,linkcolor=blue,citecolor=blue,urlcolor=blue}

\makeatletter
\@ifundefined{textcolor}{}
{%
 \definecolor{BLACK}{gray}{0}
 \definecolor{WHITE}{gray}{1}
 \definecolor{RED}{rgb}{1,0,0}
 \definecolor{GREEN}{rgb}{0,1,0}
 \definecolor{BLUE}{rgb}{0,0,1}
 \definecolor{CYAN}{cmyk}{1,0,0,0}
 \definecolor{MAGENTA}{cmyk}{0,1,0,0}
 \definecolor{YELLOW}{cmyk}{0,0,1,0}
}

\usepackage{babel}

\usepackage{color}
\usepackage{babel}
\usepackage{mathrsfs}

\@ifundefined{textcolor}{}{%
 \definecolor{BLACK}{gray}{0}
 \definecolor{WHITE}{gray}{1}
 \definecolor{RED}{rgb}{1,0,0}
 \definecolor{GREEN}{rgb}{0,1,0}
 \definecolor{BLUE}{rgb}{0,0,1}
 \definecolor{CYAN}{cmyk}{1,0,0,0}
 \definecolor{MAGENTA}{cmyk}{0,1,0,0}
 \definecolor{YELLOW}{cmyk}{0,0,1,0}
}

\usepackage{babel}
\usepackage{babel}
\usepackage{epsfig}
\usepackage{amsfonts}\usepackage{mathrsfs}\usepackage{tensor}\usepackage{psfrag}

\newcommand{\tr}{ { \rm tr}}

\usepackage{slashed}

\DeclareMathOperator{\arcsinh}{arcsinh}

\makeatother

\begin{document}

\title{Covariant Conservation Laws and the Spin Hall Effect in Dirac-Rashba
Systems}

\author{Mirco Milletarì }
\email{milletari@gmail.com}

\affiliation{Dipartimento di Matematica e Fisica, Università Roma Tre, 00146 Rome,
Italy }

\affiliation{Bioinformatics Institute, Agency for Science, Technology and Research
(A{*}STAR), Singapore 138671, Singapore}

\author{Manuel Offidani}

\affiliation{Department of Physics, University of York, York YO10 5DD, United
Kingdom}

\author{Aires Ferreira}
\email{aires.ferreira@york.ac.uk}

\affiliation{Department of Physics, University of York, York YO10 5DD, United
Kingdom}

\author{Roberto Raimondi}

\affiliation{Dipartimento di Matematica e Fisica, Università Roma Tre, 00146 Rome,
Italy }
\begin{abstract}
We present a theoretical analysis of two-dimensional Dirac-Rashba
systems in the presence of disorder and external perturbations. We
unveil a set of \emph{exact} symmetry relations (Ward identities)
that impose strong constraints on the spin dynamics of Dirac fermions
subject to proximity-induced interactions. This allows us to demonstrate
that an arbitrary dilute concentration of scalar impurities results
in the total suppression of nonequilibrium spin Hall currents when
only Rashba spin\textendash orbit coupling is present. Remarkably,
a finite spin Hall conductivity is restored when the minimal Dirac\textendash Rashba
model is supplemented with a spin\textendash valley interaction. The
Ward identities provide a systematic way to predict the emergence
of the spin Hall effect in a wider class of Dirac-Rashba systems of
experimental relevance and represent an important benchmark for testing
the validity of numerical methodologies. 
\end{abstract}
\maketitle
Systems exhibiting strong spin\textendash orbit coupling (SOC) have
received much attention because they host unique spin transport phenomena
that can be harnessed for low-power spintronics~\cite{Review_Ando_2017,Review_Fert_2016}.
The spin Hall effect (SHE)~ \cite{SHE_prediction_Dyakanov,SHE_revisted_Hirsch}
is indubitably a landmark in this novel approach; combined with its
reciprocal phenomenon (the inverse SHE), it allows all-electrical
generation, detection and manipulation of nonequilibrium spin currents
in nonmagnetic conductors~\cite{SHE1_Kato,SHE2_Wunderlich,SHE3_Saitoh_06,SHE4_Valenzuela_06}.
The exploitation of the SHE has proved fruitful for manipulation of
magnetic order via spin\textendash orbit torque at interfaces \cite{SO_NMFM_1,SO_NMFM_2,SO_NMFM_relaxation}
and has led to new discoveries, including the spin Hall magnetoresistance~\cite{SHE_magnetoresistance}.

The interest in spin\textendash orbit phenomena has been invigorated
with the recent discovery of strong Rashba splitting of two-dimensional
electron gases (2DEGs) at nonmagnetic metal surfaces and heterointerfaces~\cite{Metal_Rashba_Interactions_PhotoEmission,Metal_Rashba_Interactions,NonMagnetic_Rashba_Interactions_Review}.
Microscopically, the splitting can be understood as arising from a
potential gradient normal to the surface, $\phi(z)$, which couples
the electron spin $\mathbf{s}$ and in-plane momentum $\mathbf{p}$
i.e., in the simplest approximation, $H_{\textrm{BR}}=\alpha\,\hat{z}\cdot(\mathbf{s}\times\mathbf{p})$,
where $\alpha\propto\partial_{z}\phi$. The Rashba-Bychkov interaction
$H_{\textrm{BR}}$ (hereafter Rashba interaction) mixes orbital states
with opposite spins, leading to spin-split parabolic bands with counter-rotating
spin textures~\cite{2DEG_Rahsba_Bychkov_Effect_84}. The tangential
spin winding of Rashba states enables efficient generation of nonequilibrium
spin polarization by application of electric fields~\cite{SGE,Edelstein_effect_1,Edelstein_effect_2,Edelstein_effect_3,Edelstein_Exp_MetalHeteroInterface,Edelstein_Exp_MetalHeteroInterface_2,Edelstein_Exp_MetalHeteroInterface_1}.
Strikingly, the very helical nature of these states enforces a\emph{
vanishing} SHE in the presence of (scalar) impurity scattering \cite{SHE_2DEG_cancelation_1,SHE_2DEG_cancelation_2,SHE_2DEG_cancelation_3,Dimitrova_05,Khaetskii2006},
so that, in practice the current-induced spin polarization is not
easily accompanied by the formation of spin Hall currents~\cite{Comment-ExtrinsicSHE}.
Given the universality of the Rashba effect (also observed in ultra
thin metals \cite{UltraThin_Rashba_1,UltraThin_Rashba_2}, quantum
wells~\cite{QW_Rashba_1,QW_Rashba_2} and surfaces of topological
insulators~\cite{TI_2DEG_Rashba_at_surface,TI_2DEG_Rashba_at_surface_2,TI_2DEG_Rashba_at_surface_3}),
it is of utmost importance to understand whether the absence of SHE
is a general property of nonmagnetic surfaces with broken inversion
symmetry or, rather, a peculiarity of the 2DEG.

\begin{figure}[t]
\includegraphics[width=0.85\columnwidth]{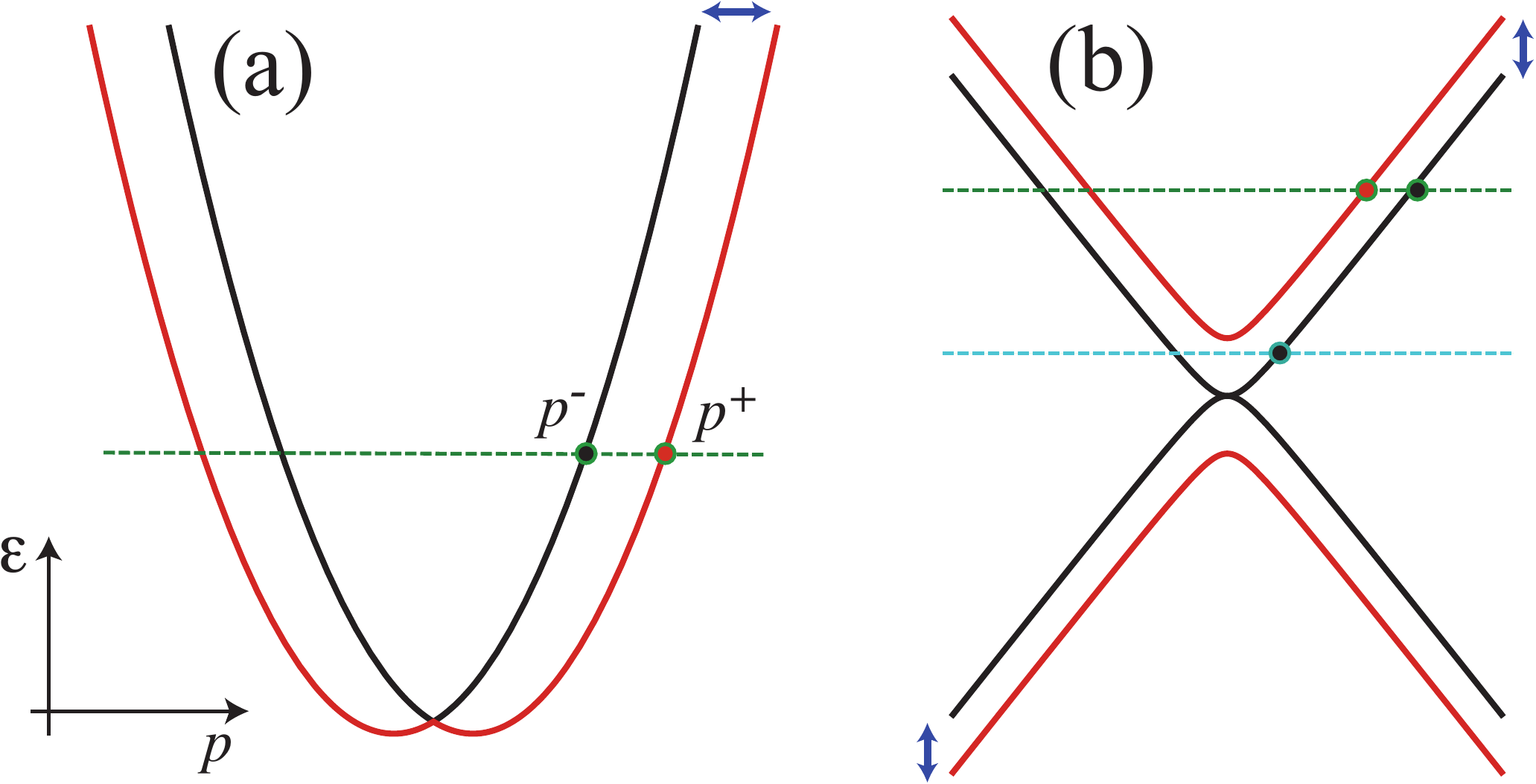} \caption{\label{fig:01}Schematic of the splitting of electronic states due
to Rashba effect in a 2DEG (a) and graphene (b). The Fermi surface
consists of two branches in a 2DEG. In graphene, for energies in the
Rashba pseudo-gap $|\epsilon|<2|\lambda|$ the Fermi surface is simply
connected. Arrows indicate the type of splitting. }
\end{figure}

The interfacial enhancement of SOC in graphene has been recently demonstrated
\cite{SOC_G_Marchenko,SOC_G_Avsar_14,SOC_G_Wang_Morpurgo_15,SOC_G_Wang_AHE_15,SOC_G_Wang_Designer_16,SOC_G_Shi_17},
making it a promising model system to explore the above issue. The
departure from the standard Rashba effect in a 2DEG can be readily
appreciated for a minimal model of graphene subject to $z\rightarrow-z$
asymmetric SOC. In the long-wavelength limit, the relevant spin\textendash orbit
interaction is obtained by replacing momentum with pseudospin operator
$\mathbf{p}\rightarrow\boldsymbol{\sigma}$ in $H_{\textrm{BR}}$~\cite{SOC_G_THEORY_Kane_Mele_05,SOG_G_THEORY_Rashba_09}.
The Hamiltonian density $\mathscr{H}=\mathscr{H}_{0}+\mathscr{H}_{\textrm{BR}}$
for the $\chi=\pm$ valley reads 
\begin{align}
\mathscr{H}=\psi_{\chi}^{\dagger}\left\{ \chi\left[-\imath\,\hbar\,v\,\sigma^{i}\,\partial_{i}\,+\lambda\,(\boldsymbol{\sigma}\times\boldsymbol{s})_{z}\right]-\epsilon\right\} \psi_{\chi},\label{eq:H_0}
\end{align}
where $v$ is the bare Fermi velocity of massless Dirac electrons,
$\lambda$ is the Rashba coupling, $\epsilon$ is the Fermi energy,
and $\sigma_{i}\,(i=1,2)$ and $s_{j}\,(j=1,2,3)$ are Pauli matrices
acting on pseudospin and spin subspace, respectively. This model possess
two noteworthy features. First, the band splitting occurs along the
energy axis {[}Fig.\,\eqref{fig:01}{]}. Secondly, the Dirac helical
spin texture is momentum dependent, i.e., $|\langle\mathbf{s}\rangle|$
is not conserved \cite{SOG_G_THEORY_Rashba_09}. Moreover, Eq.\,(\ref{eq:H_0})
admits a straighfoward generalization by adding further interactions
preserving the inherent $\textrm{SU}(2)$ spin structure, such as
a spin\textendash valley coupling. Such unique features make the Dirac\textendash Rashba
model an ideal testbed to re-examine the absence of SHE in interfaces
with spin-split states.


In this Letter, we investigate Dirac\textendash Rashba models in the
presence of disorder and external perturbations. The existence of
a covariant conservation law for the spin current\textemdash stemming
from $\textrm{SU}(2)$ gauge invariance\textemdash allows us to obtain
the analytic form of two-particle spin-current vertex functions directly
from the self energy of the Dirac fermions and show that the spin
Hall conductivity in the minimal model {[}Eq.\,(\ref{eq:H_0}){]}
is \emph{zero} for nonmagnetic disorder, \emph{irrespectively of the
Fermi level} \emph{position}. Furthermore, we show that when Eq.\,(\ref{eq:H_0})
is generalized to include additional interactions, the obtained Ward
identity imposes strong constraints on the nonequilibrium spin responses.
Remarkably, this allows us to predict what type of proximity spin\textendash orbit
interactions can lead to a robust SHE in Dirac\textendash Rashba interfaces
of experimental interest.

The suppression of SHE in 2DEGs subject to uniform Rashba interactions
occurs in the presence of an arbitrary small concentration of scalar
impurities. Formally, the disorder corrections resulting from the
resummation of ladder diagrams exactly cancel the ``clean'' spin
Hall (SH) conductivity \cite{SHE_2DEG_cancelation_1,SHE_2DEG_cancelation_2,SHE_2DEG_cancelation_3,Dimitrova_05,Khaetskii2006}.
In Ref.~\cite{Dimitrova_05} it was shown that this puzzling cancellation
has its origin in the existence of a covariant conservation law for
the spin current. For example, the spin-$y$ component satisfies 
\begin{equation}
\partial_{t}J_{0}^{y}(\mathbf{x},t)+\partial^{i}\,J_{i}^{y}(\mathbf{x},t)=-2\,\alpha\,m\,J_{y}^{z}(\mathbf{x},t),\label{eq:dim}
\end{equation}
where $J_{0}^{a}$ ($a=x,y,z$) is the spin density, $J_{i}^{a}$
is the pure spin current flowing in the $i=x,y$ direction, $m$ is
the effective electron mass, and $\alpha$ is the Rashba parameter.
The main difference with respect to the charge continuity equation
originates from the non Abelian nature of spin, which results in the
additional contribution on the right hand side. Equation~\eqref{eq:dim}
suggests that in the steady state of a homogeneous system, $J_{y}^{z}$
is zero irrespectively of the underlying relaxation mechanism. Below
we show that, albeit the drastically different nature of electronic
states in the Dirac\textendash Rashba model {[}Fig.~\eqref{fig:01}{]},
a similar covariant conservation law exists, and discuss its consequences.


\emph{Conservation laws I}.\textemdash A peculiarity of Dirac theories
is the possible existence of quantum anomalies due to the joint effect
of an infinite Dirac sea of filled electron states and an external
field~\cite{frad,bert}. Let us consider the minimal coupling of
Eq.\,(\ref{eq:H_0}) to a $U(1)$ gauge field $A_{\mu}\equiv(A_{0},A_{i})$
within a Minkowsky metric. To simplify notation, we take $\chi=+$
and omit this index hereafter. We also use natural units ($\hslash\equiv1\equiv e$)
and the compact notation $\partial_{\mu}\equiv(\partial_{t},\partial_{i})$
with summation over dummy indices. The Dirac spin and charge currents
are, respectively, $J_{\mu}^{a}(x)=\psi^{\dagger}(x)\,s^{a}v_{\mu}/2\,\psi(x)$
and $J_{\mu}(x)=\psi^{\dagger}(x)v_{\mu}\psi(x)$, where $v^{\mu}=(1,v\,\boldsymbol{\sigma})$
and $x\equiv(t,\mathbf{x})$. The Heisenberg equation of motion for
the spin density reads

\begin{align}
\partial^{\mu}J_{\mu}^{a}(x)=-\frac{2\lambda}{v}\epsilon_{bc}^{a}\,\epsilon^{bl}J_{l}^{c}(x)+\imath\int dy[J_{0}^{a}(x),J_{\mu}(y)]A^{\mu}(y),\label{eq:Heis}
\end{align}
where $\epsilon^{bl}$\,($\epsilon_{bc}^{a}$) is the Levi-Civita
symbol of second (third) rank. The term on the left hand side and
the first on the right result from the commutator of $J_{0}^{a}$
respectively with the kinetic and the Rashba term and give a contribution
identical to the one found in the 2DEG upon identification of $m\to1/v$,
c.f.\,\,Eq.\,\eqref{eq:dim}. Both terms can be combined as the
covariant derivative $D_{\mu}{\cal O}^{a}=\partial_{\mu}{\cal O}^{a}+2\,\epsilon_{bc}^{a}\,{\cal A}_{\mu}^{b}{\cal O}^{c}$,
where ${\cal A}_{0}^{a}=0$, ${\cal A}_{i}^{a}=-\,\lambda/v\,\epsilon^{ai}$
is a SOC-induced, homogeneous gauge field. Hence, in the absence of
an external field, Eq.\,(\ref{eq:Heis}) acquires the form of a covariant
conservation law for the spin density $D^{\mu}\,J_{\mu}^{a}=0$. The
current commutator in the last term (Schwinger term) defines the anomaly.
A careful analysis shows however that despite the Dirac nature of
the theory, the commutator is identically zero \textendash{} see supplemental
material~\cite{suppl}; therefore, the argument of Ref.~\cite{Dimitrova_05}
implies a vanishing SHE in the Dirac\textendash Rashba model. At first
sight this result contradicts the claims of Ref.~\cite{spin_current_G_Dyrdal},
where the SH conductivity was evaluated using linear response theory
$\sigma_{\textrm{SH}}=\lim_{\omega\to0}\lim_{q\to0}\,\Theta_{yx}^{z}(\mathbf{q},\omega)/i\omega$,
with the response function $\Theta_{yx}^{z}$ taken in the disorder-free
approximation. Using the Matsubara propagator given in~\cite{suppl}
we find %
\begin{equation}
\sigma_{\textrm{SH}}=-\frac{\epsilon}{16\pi\lambda}\left[\frac{2\lambda+\epsilon}{\epsilon+\lambda}+\theta(\epsilon-2\lambda)\frac{2\lambda-\epsilon}{\epsilon-\lambda}\right]\;,\label{eq:clean}
\end{equation}
in agreement with Ref.~\cite{spin_current_G_Dyrdal}. Here $\theta(.)$
is the Heaviside step function and we assumed $\epsilon,\lambda>0$.
The apparent contradiction is resolved by recalling that, without
disorder, there is no true stationary state. In the following we show
that Eq.~\eqref{eq:clean} misses on important physics related to
scattering-induced relaxation that leads to $\sigma_{\textrm{SH}}=0$.

\emph{Conservation laws II: disorder effects}.\textemdash Broadly
speaking, the Fermi surface contribution to $\sigma_{\textrm{SH}}$
is dominated by incoherent multiple scattering off impurities, which
can be viewed as a series of skew scattering and side jump events
\cite{SHE/AHE_theory_1_Bruno,SHE_theory_2_Synitsin,SHE_theory_3_Mirco}.
To determine how such effects change the above picture, we add to
the bare Hamiltonian~\eqref{eq:H_0} a random scalar potential $V(\mathbf{x})$,
which we will assume to be Gaussian distributed with zero mean: $\langle V(\mathbf{x})V(\mathbf{x}')\rangle=n_{i}\,\alpha_{0}^{2}\,\delta(\mathbf{x}-\mathbf{x}')$,
where $n_{i}$ is the impurity areal density and $\alpha_{0}$ parametrizes
the potential strength. This approximation is accurate in the limit
of weak potential scattering provided cross sections are right\textendash left
symmetric (see below). We note that short-range impurities lead to
scattering potentials that are off-diagonal in both sublattice and
valley spaces. The intervalley scattering produced by such matrix
disorder affects the charge conductivity $\sigma_{xx}$ \cite{G_transport_theory_Peres},
but it does not change the covariant conservation law for the spin
current.

Disorder enters the evaluation of response functions both in the propagator
(as a self-energy) and the interaction vertex~\cite{DicastroRaimondi}.
These two quantities are \emph{not} independent of each other but
they are related by Ward identities (WIs); these relations are the
key to establish gauge invariance in quantum electrodynamics at a
non-perturbative level~\cite{bert}. Remarkably, we find that the
non-Abelian WI associated to the spin current vertex completely determines
the spin current $J_{i}^{z}$ in the dc limit and therefore it can
be used to directly evaluate the SH conductivity. To see this, consider
the three-legged spin vertex function $\Lambda_{\mu}^{y}(x,x',x'')=\langle T_{\tau}\,J_{\mu}^{y}(x)\,\psi(x')\,\psi^{\dagger}(x'')\rangle$,
where ``$T_{\tau}$'' stands for the imaginary time ordering operator.
Moving to frequency-momentum space, we perform analytic continuation
$\imath\,\omega_{n}\to\omega+\imath\,{\rm sign}(\omega)\,0^{+}$,
where $\omega_{n}$ are fermionic Matsubara frequencies. Vertex corrections
appear perturbatively as a series of impurity lines ladder diagrams,
where only combinations of Green's functions having poles on opposite
sides of the real axis contribute to the renormalization of the vertex~\cite{DicastroRaimondi}.
In this way, by projecting the vertex function $\Lambda_{\mu}^{y}$
in the retarded (R) \textemdash advanced (A) sector, we find 
\begin{equation}
q^{\mu}\Lambda_{\mu}^{y}=-\imath\,\frac{2\lambda}{v}\,\Lambda_{y}^{z}+\frac{1}{2}\left(s_{y}\,\mathcal{G}_{k+q}^{R}(\epsilon)-\mathcal{G}_{k}^{A}(\epsilon)\,s_{y}\right),\label{eq:WI1}
\end{equation}
where $k$ and $q$ are three-vectors. The disorder averaged Green's
function ($a=A,R$) formally reads $\mathcal{G}_{k}^{a}(\epsilon)=\left[k_{0}-H-\Sigma^{a}(\epsilon)\right]^{-1}$,
where $H$ is given by the first quantization form of Eq.~\eqref{eq:H_0}
and $\Sigma^{a}(\epsilon)$ is the disorder induced self energy (see
SM~\cite{suppl} for an explicit form). Owing to the non-Abelian
nature of the WI, taking the dc ($q\to0$) limit in Eq.\,(\ref{eq:WI1})
completely determines the effective vertex. The final step consists
in recasting $\Lambda_{y}^{z}$ in terms of the truncated vertex,
$\Lambda_{y}^{z}=\mathcal{G}^{A}\,\tilde{j}_{y}^{z}\,\mathcal{G}^{R}$~\cite{Comment_truncated},
as appearing in the Kubo formula. After algebraic manipulations, we
arrive at the important intermediate result 
\begin{equation}
\tilde{j}_{y}^{z}=-\imath\frac{v}{4\lambda}\left\{ [s_{y},\tilde{H}]_{-}+\imath\thinspace[s_{y},{\rm Im}\,\Sigma^{R}(\epsilon)]_{+}\right\} \thinspace,\label{eq:WI2}
\end{equation}
where $\pm$ stands for the (anti-)commutator and $\tilde{H}=H+\textrm{Re}\,\Sigma$
is the Hamiltonian renormalized by the real part of the self energy.
This result provides an \emph{exact} relation between the truncated
spin current vertex and the self energy, and as such, it is independent
of the particular approximation scheme used to evaluate disorder effects.
Within the Gaussian approximation, we find $-{\rm Im}\,\Sigma^{R}(\epsilon)=1/(2\tau)[1+\theta(2\lambda-\epsilon)\lambda/2\epsilon]\sigma_{0}s_{0}+\theta(2\lambda-\epsilon)[\lambda/(2\tau\epsilon)\,\sigma_{3}s_{3}-1/(8\tau)\,(\boldsymbol{\sigma}\times\boldsymbol{s})_{z}]$,
where $1/2\tau=n_{i}\,\epsilon\,\alpha_{0}^{2}/4\,v^{2}$ is the quasiparticle
broadening. Using the expression of the self energy in Eq.\,\eqref{eq:WI2},
we arrive at 
\begin{equation}
\tilde{j}_{y}^{z}=\frac{v}{2}\times\begin{cases}
\sigma_{y}s_{z}-\frac{1}{2\lambda\tau}\sigma_{0}s_{y} & \,,\epsilon>2\lambda\\
\sigma_{y}s_{z}-\frac{1}{4\lambda\,\tau}\left(1+\frac{\lambda}{\epsilon}\right)\sigma_{0}s_{y}\\
+\frac{1}{8\lambda\tau}\sigma_{x}s_{0}+\frac{1}{4\pi\,\tau\lambda}\sigma_{z}s_{x}. & \,,\epsilon\le2\lambda
\end{cases}\label{eq:WI4}
\end{equation}
The first term is just the bare spin current vertex $j_{y}^{z}=\frac{v}{2}\sigma_{y}s_{z}$,
while, for $\epsilon>2\lambda$, the second term, generated by the
disorder, is the bare spin density vertex $\sigma_{0}s_{y}/2$ apart
from the factor $-v/2\,\lambda\,\tau$. This shows that the parameter
$\lambda\tau$ plays a fundamental role in determining the importance
of disorder. At first sight one could be tempted to think that within
the weak disorder limit ($\epsilon\,\tau\gg1$) and for strong SOC
($\lambda\,\tau\gg1$), all disorder corrections can be neglected.
However, it turns out that the spin polarization response is of order
$\lambda\,\tau$ (see below), whereas the bare spin current response,
due to the first term in Eq.\,(\ref{eq:WI4}) is of order $(\lambda\,\tau)^{0}$.
Hence, the two terms are of the same order irrespective of the disorder
strength. Similar considerations apply also for $\epsilon<2\lambda$.

\emph{SHE evaluation using the WI}.\textemdash We start by computing
the Fermi surface contribution 
\begin{align}
\sigma_{\textrm{SH}}^{\textrm{I}} & =\frac{1}{2\pi}\int\frac{d\mathbf{k}}{(2\pi)^{2}}\,\textrm{tr}\,\left[\tilde{j}_{y}^{z}\,\mathcal{G}_{\mathbf{k}}^{R}(\epsilon)\,v_{x}\,\mathcal{G}_{\mathbf{k}}^{A}(\epsilon)\right]\nonumber \\
 & =\bar{\sigma}_{\textrm{SH}}+\bar{\sigma}_{SG}+\bar{\sigma}_{xx}+\bar{\sigma}_{zx}\,\label{eq:SH1_def}
\end{align}
where $v_{x}=v\,\sigma_{x}s_{0}$ is the bare charge current vertex.
Moreover, $\bar{\sigma}_{\textrm{SH}}$, $\bar{\sigma}_{\textrm{SG}}$,
$\bar{\sigma}_{xx}$, and $\bar{\sigma}_{zx}$ are the conductivity
``bubbles'' corresponding to the various terms in Eq.\,\eqref{eq:WI4},
respectively, a spin Hall ($\sigma_{y}s_{z})$, spin galvanic (SG)
($\sigma_{0}s_{y}$), longitudinal $(\sigma_{x}s_{0}$) and ``staggered''
($\sigma_{z}s_{x}$) conductivities. Outside the pseudo-gap, where
the Fermi surface splits into two branches {[}Fig.\,(\ref{fig:01}){]},
we find $\bar{\sigma}_{xx}=\bar{\sigma}_{zx}=0$ and $\bar{\sigma}_{\textrm{SH}}=-\bar{\sigma}_{\textrm{SG}}$,
where 
\begin{align}
\bar{\sigma}_{\textrm{SH}}=-\frac{1}{8\pi}\left(\frac{\epsilon^{2}}{\epsilon^{2}-\lambda^{2}}-\frac{1}{1+4\,\lambda^{2}\,\tau^{2}}\right)\,,\label{sh:bubble}
\end{align}
and thus the type I contribution to the SH conductivity is zero, $\sigma_{\textrm{SH}}^{\textrm{I}}=0$.
This result deserves few comments: First, in the $\lambda\,\tau\gg1$
limit, one recovers Eq.~(\ref{eq:clean}). Second, the ``empty bubble''
SH conductivity ($\bar{\sigma}_{\textrm{SH}}$) is precisely counteracted
by the corresponding ``empty bubble'' for spin density-charge current
response function ($\bar{\sigma}_{\textrm{SG}}$) \cite{Comment_SG_bubble}.
This means that the absence of SHE is linked to the onset of a current-induced,
in-plane spin polarization known as the inverse spin galvanic effect~\cite{SGE,Edelstein_effect_1,Edelstein_effect_2}.
The remaining (type II) contribution 
\begin{equation}
\sigma_{\textrm{SH}}^{\textrm{II}}=\frac{-1}{2\pi}\int\frac{d\mathbf{k}}{(2\pi)^{2}}\int_{-\infty}^{0}dk_{0}\,\textrm{Re}\,\textrm{tr}\left[\mathcal{G}_{k}^{R}(\epsilon)\,j_{y}^{z}\,\overleftrightarrow{\partial_{k_{0}}}\,\mathcal{G}_{k}^{R}(\epsilon)\,v_{x}\right]\,,\label{eq:Streda}
\end{equation}
accounts for processes away from the Fermi surface~\cite{Streda_}.
Explicit evaluation shows that $\sigma_{\textrm{SH}}^{\textrm{II}}=0$
and thus $\sigma_{\textrm{SH}}=\sigma_{\textrm{SH}}^{\textrm{I}}+\sigma_{\textrm{SH}}^{\textrm{II}}$
is zero, in agreement with our earlier argument viz., Eqs.\,(\ref{eq:dim})-(\ref{eq:Heis}).
Interestingly, in the 2DEG\textendash Rashba model, the type II term
is only zero in the formal limit $\epsilon\tau\rightarrow\infty$
and can attain large values for $\lambda\,\tau\approx1$ \cite{Grimaldi06}.
The exact vanishing of the off-Fermi surface contribution is a \textit{unique
feature} of the Dirac theory. We now move gears to the regime $\epsilon<2\lambda$,
where only one subband is occupied. We note that this regime has no
analogue in the 2DEG model, for which the Fermi surface always consists
of two disconnected rings {[}Fig.\,(\ref{fig:01}){]}. The mechanism
leading to $\sigma_{\textrm{SH}}=0$ is thus far from obvious. To
investigate this issue, we evaluate the Fermi surface contribution
making use of the WI {[}see Eq.\,(\ref{eq:WI4}){]} and the type
II contribution using~Eq.\,(\ref{eq:Streda}). After a lengthy calculation,
we find for both contributions 
\begin{equation}
\sigma_{\textrm{SH}}^{\textrm{I}}=\frac{1}{16\pi}\frac{\epsilon}{\lambda}\quad,\quad\sigma_{\textrm{SH}}^{\textrm{II}}=-\sigma_{\textrm{SH}}^{\textrm{I}}\,.\label{eq:type_I_II}
\end{equation}
so that $\sigma_{\textrm{SH}}=0$. Note that since $\sigma_{\textrm{SH}}^{\textrm{I}}$
is of order $\tau^{0}$, we can evaluate the type II contribution
{[}Eq.~\eqref{eq:Streda}{]} directly in the absence of disorder.
The suppression of the SHE in the regime $0<\epsilon<2\lambda$ therefore
results from a compensation between scattering corrections to the
``clean'' SH conductivity and off-Fermi surface processes.

\begin{figure}[t!]
\includegraphics[width=0.6\columnwidth]{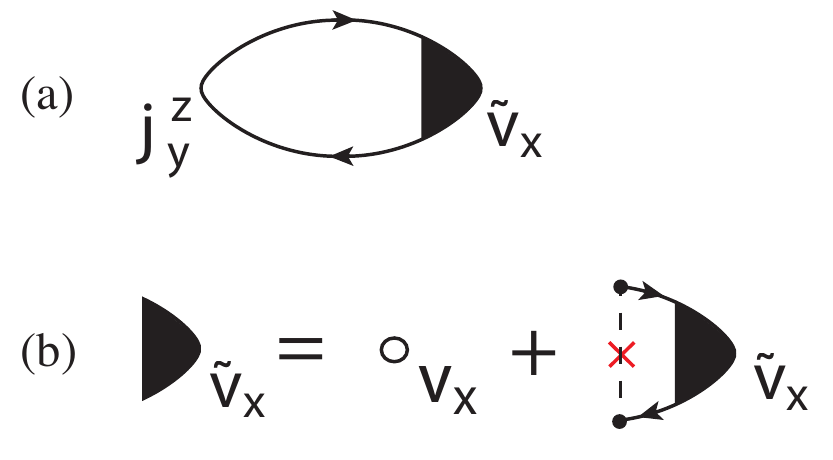} \caption{\label{fig:02}Feynman diagrams for (a) dressed SH conductivity. (b)
Charge vertex renormalization. The empty dot represents the bare charge
vertex while the red $x$ and the black dots represent respectively
impurity density and scattering potential insertions.}
\end{figure}

\emph{Diagrammatic evaluation}.\textemdash We now show the consistency
of our results with a standard diagrammatic evaluation. The renormalized
charge current vertex satisfies the following Bethe-Salpeter coupled
equations {[}see Fig.~\eqref{fig:02}{]} 

\begin{align}
\tilde{v}_{x,\mu a} & =v\,\delta_{\mu1}\,\delta_{a0}+T\indices{_{\mu a\rho d}^{\nu b\lambda c}}\,I_{\nu b\lambda c}\,\tilde{v}_{x}^{\rho d}\,,\label{eq:vert1}\\
T\indices{_{\mu a\rho d}^{\nu b\lambda c}} & =\textrm{tr}\left[\sigma_{\mu}\,s_{a}\,\sigma_{\nu}\,s_{b}\,\sigma_{\rho}\,s_{d}\,\sigma_{\lambda}\,s_{c}\right]\,,\label{eq:vert2}\\
I_{\nu b\lambda c} & =\frac{n_{i}\,\alpha_{0}^{2}}{4}\,\int\frac{d\mathbf{k}}{(2\pi)^{2}}\,\mathcal{G}_{\mathbf{k},\nu b}^{R}(\epsilon)\,\mathcal{G}_{\mathbf{k},\lambda c}^{A}(\epsilon).\label{eq:vert3}
\end{align}
In principle, $I$ spans the entire Clifford Algebra. However, not
all matrix elements contribute to the renormalization of the charge
vertex. It is convenient to consider the effect of a single impurity
density insertion, for which the vertex has the structure: $\bar{v}_{x}=\delta v_{10}\,\sigma_{1}\,s_{0}+\delta v_{23}\,\sigma_{2}\,s_{3}+\delta v_{02}\,\sigma_{0}\,s_{2}+\delta v_{31}\,\sigma_{3}\,s_{1}$,
with $\delta v_{ij}$ some non-zero matrix elements. This result suggests
the form of the ansatz for $\tilde{v}_{x}$ to use in Eq.~\eqref{eq:vert1}.
Since no new matrix element is generated in this procedure, the ansatz
closes the system. In addition to the renormalized charge vertex $\tilde{v}_{x}^{10}$,
we find that disorder induces an effective SH ($\tilde{v}_{x}^{23}$),
spin galvanic ($\tilde{v}_{x}^{02}$) and ``staggered'' ($\tilde{v}_{x}^{31}$)
interaction. Their explicit form reads (for $\epsilon>2\lambda$):
$\tilde{v}_{x}^{10}=2\,v$, $\tilde{v}_{x}^{02}=-2\,v(\lambda/\epsilon)$,
$\tilde{v}_{x}^{31}=0$ and $\tilde{v}_{x}^{23}=0$. In order to evaluate
the SH conductivity we use now Eq.~\eqref{eq:SH1_def}, with the
ladder series now included in the charge vertex (i.e. $\tilde{j}_{y}^{z}\to j_{y}^{z}$
and $v_{x}\to\tilde{v}_{x}$). Using Eq.~\eqref{eq:vert1}, it is
now easy to relate the renormalized vertex directly to the SH and
Drude conductivity %
\begin{align}
\sigma_{\textrm{SH}} & =\frac{1}{2\,\pi}\left(\frac{2\,v}{n_{i}\,\alpha_{0}^{2}}\right)\tilde{v}_{x}^{23}=0,\\
\sigma_{xx} & =\frac{1}{2\,\pi}\left(\frac{4\,v}{n_{i}\,\alpha_{0}^{2}}\right)(\tilde{v}_{x}^{10}-v)=\frac{2\,\epsilon\,\tau}{\pi}.\label{eq:16}
\end{align}
{\emph{Discussions}.\textemdash We mentioned earlier that higher-order
scattering contributions to the self energy (and ladder series) could
generate important corrections. This happens when impurities in the
system lead to skew scattering. In the 2DEG, it is well known that
skew scattering is absent (unless other ingredients, such as spin\textendash orbit
active impurities are considered). The absence of skewness has in
fact an intuitive explanation: the spin of Rashba eigenstates is locked
in-plane, so that in a given scattering event quasiparticles cannot
distinguish left and right. The same picture holds in the Dirac\textendash Rashba
model and so here too there should be no skewness. We verified this
by means of the self-consistent diagrammatic approach introduced in
Ref.~\cite{SHE_theory_3_Mirco} together with the WI {[}Eq.\,(\ref{eq:WI2}){]}.

The formalism developed in this Letter also allows to predict the
behaviour of more complicated systems. For instance, it is easy to
see that a non-zero SH conductivity emerges when adding suitable interactions
to Eq.\,(\ref{eq:H_0}), altering the covariant conservation law
expressed in Eq.\,(\ref{eq:Heis}) and hence the WI {[}Eq.\,(\ref{eq:WI2}){]}.
For example, let us consider a spin\textendash valley interaction
of the form ${\cal A}_{0}^{a}=\chi\lambda^{\prime}\,\delta_{az}$
with $\lambda^{\prime}$ a constant. This interaction generates in
Eq.\,(\ref{eq:Heis}) a new term proportional to $\langle s_{x}^{\chi}\rangle$,
where $\langle s_{x}^{\chi}\rangle$ is the nonequilibrium average
of the $\hat{x}$-spin polarization at a given valley. Taking the
steady state of a homogeneous system, we find an exact relation between
the spin Hall current and the difference between the nonequilibrium
spin density at the two inequivalent valleys, namely: %
\begin{equation}
\langle J_{y}^{z}\rangle=v\,\frac{\lambda^{\prime}}{\lambda}\left(\langle s_{x}^{\chi=1}\rangle-\langle s_{x}^{\chi=-1}\rangle\right).\label{eq:prediction}
\end{equation}
This suggests that SHE can emerge provided there is a mechanism to
generate $\langle s_{x}^{\chi}\rangle\neq0$ with opposite signs for
$\chi=\pm1$. A strong candidate is skew scattering. In principle,
skewness is now allowed since the spin\textendash valley interaction
takes the spin of bare eigenstates out of the plane. We have computed
both (non-vanishing) sides of Eq.~(\ref{eq:prediction}) diagrammatically
and verified that the identity holds at all orders in the scattering
potential strength (not shown). This is a significant finding since
the spin\textendash valley coupling $\lambda^{\prime}$ can attain
sizable values in graphene with proximity SOC~\cite{SOC_G_theory_Gmitra16,SOC_G_theory_Kochan_17}.
The possibility to have skew scattering exclusively driven by SOC
in the band structure appears to be a unique feature of Dirac systems.

In this context, we note in passing that random spatial fluctuations
in the Rashba coupling (e.g., due to corrugations) provide an alternative
source of SHE \cite{SHE_RandomSOC}. The skew scattering contribution
discussed above is dominant in clean samples due to its characteristic
scaling ($n_{i}^{-1}$ opposed to $n_{i}^{0}$ in the random mechanism)
and the relatively small size of the fluctuations expected for atomically-flat
interfaces.

Our work constitutes a major step towards a unified theory of spin
and charge dynamics for Dirac-Rashba models in generic non-stationary
conditions. Real-space methodologies for numerical evaluation of transverse
conductivities have been recently proposed~\cite{KPM_Rappoport_15,WavePacket_Ortmann_15},
which can help tackling more complex scenarios. The exact symmetry
relations presented here provide a stringent test for real-space numerical
approaches, for which the achievable energy resolutions still represent
a major limiting factor.\\

Data availability statement (EPSRC).\textendash No new data were created
during this study. \\

We would like to thank G. Vignale, M.A. Cazalilla, C. Huang , and
J. Song for useful discussions. M. M. thanks C. Verma for his hospitality
at the Bioinfomatics Institute in Singapore. A.F. gratefully acknowledges
the financial support from the Royal Society (U.K.) throug h a Royal
Society University Research Fellowship. R.R. acknowl- edg es the hospitality
of Centre for Advanced 2D Materials (CA2DM) at National University
of Singapore (NUS) under Grant No. R-723-000-009-281 (GL 769105).
M. O. and A. F. acknowledg e funding from EPSRC (Grant No. EP/N004817/1).


\newpage{}

\begin{widetext} 
\begin{center}
\textbf{\large{}{}{}{}{}{}{}{{Supplemental Material}}{}{}}{\large{}{}{}{}{}{}{}{{{
} \setcounter{equation}{0}}}} {\large{}{}{}{}{}{}{}{{\setcounter{figure}{0}}}} 
\par\end{center}

In this supplemental material we provide explicit expressions for
the propagators in the presence of Rashba spin orbit coupling (both
in the clean and disordered case) used in the main text. We provide
an explicit proof of the absence of anomalous commutators (Schwinger
terms) in the continuity equation for the spin current and finally
give some details on the evaluation of the SH conductivity. 

\section{Spinor propagator with Rashba SOC}

\label{ch:SpinP} Due to the breaking of the spin rotational symmetry
$SU_{s}(2)$ in the presence of the Rashba term, the fermionic propagator
is not easily invertible. Here we show how to obtain the general expression
of the propagator both in the clean and in the disordered case. The
Dirac-Rashba Hamiltonian density in $\chi=\pm$ valley is given in
Eq.~(1) of the main text. In momentum space it reads 
\begin{equation}
\mathscr{H}_{\chi}=\psi_{\chi}^{\dagger}\,\chi\Big\{ v\,\boldsymbol{\sigma}\cdot\mathbf{k}+\lambda\,(\sigma_{1}\,s_{2}-\sigma_{2}\,s_{1})\Big\}\,\psi_{\chi}\label{eq:hkDR}
\end{equation}
with eigenvalues 
\begin{equation}
E_{ab,\chi}(k)=\chi\,(a\,\lambda+b\sqrt{(v\,k)^{2}+\lambda^{2}}),\label{eq:eigenvalues}
\end{equation}
where $b=\pm1$ indexes the positive/negative energy bands and $a=\pm1$
indexes the ``helicity'' sub-band~\cite{spincomm}. It is convenient
to introduce the angle $\theta_{a}=\arcsinh(-a\,\lambda/vk)$ in order
to rewrite the eigenstates in compact form as 
\begin{equation}
\Phi_{ab}=\frac{1}{2\sqrt{\cosh\theta_{a}}}\left(\begin{array}{c}
-\imath\,a\,b\,e^{-\imath\phi}\,e^{b\theta_{a}/2}\\
e^{-b\,\theta_{a}/2}\\
-\imath\,a\,e^{-b\,\theta_{a}/2}\\
b\,e^{\imath\,\phi}\,e^{b\,\theta_{a}/2}
\end{array}\right),\label{ap:Est}
\end{equation}
where $\phi$ is the angle between $k_{x}$ and $k_{y}$. The projector
over the basis of the energy eigenstates of the Hamiltonian \eqref{eq:hkDR}
is then $P_{ab}=|\Phi_{ab}\rangle\langle\Phi_{ab}|$. For computational
purpose, it is convenient to expand again the projector over the $SU_{p}(2)\times SU_{s}(2)$
spinor representation 
\begin{align}
P_{ab} & =\frac{1}{4}\Big\{(\sigma_{0}s_{0})+(\sigma_{3}s_{3})\,b\,\tanh\theta_{a}+(\sigma^{i}s_{0})\,\hat{k}_{i}\,\frac{b}{\cosh\theta_{a}}+\,\frac{(\sigma_{1}s_{2})\,a}{2\,\cosh\theta_{a}}(e^{b\,\theta_{a}}\,\cos2\phi+e^{-b\,\theta_{a}})+\frac{(\sigma_{2}\,s_{2})\,a}{2\,\cosh\theta_{a}}\,e^{b\,\theta_{a}}\,\sin2\phi\nonumber \\
 & +(\sigma_{0}s_{2})\frac{b\,a}{\cosh\theta_{a}}\,\cos\phi+\,\frac{(\sigma_{2}s_{1})\,a}{2\,\cosh\theta_{a}}(e^{b\,\theta_{a}}\,\cos2\phi-e^{-b\,\theta_{a}})-\frac{(\sigma_{1}\,s_{1})\,a}{2\,\cosh\theta_{a}}\,e^{b\,\theta_{a}}\,\sin2\phi+(\sigma_{0}s_{1})\frac{b\,a}{\cosh\theta_{a}}\,\sin\phi\Big\}
\end{align}
Note that we use the notation $k\equiv|\mathbf{k}|$ and $\hat{k}_{i}=k_{i}/k$.
Using $P_{ab}$, we can rewrite the clean Matsubara propagator as
\begin{equation}
G(\mathbf{k},\imath\,\nu_{n})=\sum_{ab,\chi}\frac{P_{ab}}{\epsilon+\imath\,\nu_{n}-E_{ab,\chi}},\label{ap:Mats}
\end{equation}
where $\nu_{n}$ are fermionic Matsubara frequencies. The real time,
disorder averaged propagator in the Retarded (R)/ Advanced (A) sector
reads instead 
\begin{equation}
\mathcal{G}^{R/A}(\mathbf{k})=\sum_{ab,\chi}\frac{P_{ab}}{\epsilon-\bar{E}_{ab,\chi}\pm\,\imath\,n_{i}\,\eta(\bar{E}_{ab,\chi})},\label{ap:DisPr}
\end{equation}
where $n_{i}$ is the density of impurities, $\bar{E}_{ab,\chi}$
are the energy eigenvalues after disorder average. Finally, $\eta(\bar{E}_{ab})$
is an energy dependent disorder broadening, whose value depends on
whether the Fermi level is inside or outside the Rashba pseudogap
(see main text). Below we provide the explicit expression for the
matrix elements of these two propagators. 

\section{Matsubara Propagator}

\label{sec:Matsubara} Here we list the matrix elements of the $\chi-$\textit{valley}
propagator. We expand the propagator over the spinor basis as $G_{\chi}(\mathbf{k},\imath\,\nu_{n})=\sigma^{\mu}s^{\nu}\,G_{\chi,\mu\nu}(\mathbf{k},\imath\,\nu)$,
where summation over repeated indices is understood. The only non-zero
matrix elements are 
\begin{align}
G_{\chi,00}(\mathbf{k},\imath\,\nu_{n}) & =-\frac{1}{2}\Big\{(\epsilon+\imath\,\nu_{n}+\chi\,\lambda)\,L_{1,\chi}(k,\imath\,\nu_{n})+(\epsilon+\imath\,\nu_{n}-\chi\,\lambda)\,L_{2,\chi}(k,\imath\,\nu_{n})\Big\}\label{ap:MatsList}\\
G_{\chi,i0}(\mathbf{k},\imath\,\nu_{n}) & =-\frac{\chi\,v\,k_{i}}{2}\Big\{ L_{1,\chi}(k,\imath\,\nu_{n})+L_{2,\chi}(k,\imath\,\nu_{n})\Big\}\\
G_{\chi,12}(\mathbf{k},\imath\,\nu_{n}) & =\frac{\cos2\phi}{4}\Big\{(\epsilon+\imath\,\nu_{n}+2\chi\,\lambda)\,L_{1,\chi}(k,\imath\,\nu_{n})-(\epsilon+\imath\,\nu_{n}-2\chi\,\lambda)\,L_{2,\chi}(k,\imath\,\nu_{n})\Big\}\\
 & +\frac{1}{4}(\epsilon+\imath\,\nu_{n})\Big\{ L_{1,\chi}(k,\imath\,\nu_{n})-L_{2,\chi}(k,\imath\,\nu_{n})\Big\}\nonumber \\
G_{\chi,21}(\mathbf{k},\imath\,\nu_{n}) & =\frac{\cos2\phi}{4}\Big\{(\epsilon+\imath\,\nu_{n}+2\chi\,\lambda)\,L_{1,\chi}(k,\imath\,\nu_{n})-(\epsilon+\imath\,\nu_{n}-2\chi\,\lambda)\,L_{2,\chi}(k,\imath\,\nu_{n})\Big\}\\
 & -\frac{1}{4}(\epsilon+\imath\,\nu_{n})\Big\{ L_{1,\chi}(k,\imath\,\nu_{n})-L_{2,\chi}(k,\imath\,\nu_{n})\Big\}\nonumber \\
G_{\chi,11}(\mathbf{k},\imath\,\nu_{n}) & =-\frac{\sin2\phi}{4}\Big\{(\epsilon+\imath\,\nu_{n}+2\chi\,\lambda)\,L_{1,\chi}(k,\imath\,\nu_{n})-(\epsilon+\imath\,\nu_{n}-2\chi\,\lambda)\,L_{2,\chi}(k,\imath\,\nu_{n})\Big\}\\
G_{\chi,22}(\mathbf{k},\imath\,\nu_{n}) & =-G_{\chi,11}(k,\imath\,\nu_{n})\\
G_{\chi,01}(\mathbf{k},\imath\,\nu_{n}) & =-\frac{\chi\,v\,k\,\sin\phi}{2}\Big\{ L_{1,\chi}(k,\imath\,\nu_{n})-L_{2,\chi}(k,\imath\,\nu_{n})\Big\}\\
G_{\chi,02}(\mathbf{k},\imath\,\nu_{n}) & =\frac{\chi\,v\,k\,\cos\phi}{2}\Big\{ L_{1,\chi}(k,\imath\,\nu_{n})-L_{2,\chi}(k,\imath\,\nu_{n})\Big\}\\
G_{\chi,33}(\mathbf{k},\imath\,\nu_{n}) & =-\frac{\chi\,\lambda}{2}\Big\{ L_{1,\chi}(k,\imath\,\nu_{n})-L_{2,\chi}(k,\imath\,\nu_{n})\Big\}.
\end{align}
We have defined the two ``Kernels'' 
\begin{align}
L_{1,\chi}(k,\imath\,\nu_{n}) & =\frac{1}{v^{2}\,k^{2}-(\epsilon+\imath\,\nu_{n})(\epsilon+\imath\,\nu_{n}+2\,\chi\,\lambda)}\label{ap:kern}\\
L_{2,\chi}(k,\imath\,\nu_{n}) & =\frac{1}{v^{2}\,k^{2}-(\epsilon+\imath\,\nu_{n})(\epsilon+\imath\,\nu_{n}-2\,\chi\,\lambda)}
\end{align}
Note that under valley exchange, the two Kernels are also interchanged,
i.e $L_{1,-1}=L_{2,1}$ and vice versa.

\section{Disorder averaged propagator}

\label{sec:disp} Here we list the matrix elements of the real time,
disorder averaged propagator in the Retarded (R)/ Advanced (A) sector.
We use the notation $\mathcal{G}_{\chi}(\mathbf{k})=\sigma^{\mu}s^{\nu}\,\mathcal{G}_{\chi,\mu\nu}(\mathbf{k})$,
where the Green's functions are now evaluated at zero real frequencies.
We first define the following notation for the Heaviside step function:
$\theta_{1/2}=\theta(\epsilon\mp2\lambda)$. In the gaussian approximation,
we define the following parameters 
\begin{align}
m & =\frac{\lambda}{4\pi\,\tau\,\epsilon}\log\frac{|\epsilon-2\lambda|}{\epsilon+2\lambda}\label{eq:disp}\\
\delta m & =\frac{\lambda}{4\,\tau\,\epsilon}(\theta_{1}-\theta_{2}),\\
\eta & =\frac{1}{4\,\tau}(\theta_{1}+\theta_{2})+\frac{\lambda}{4\,\tau\,\epsilon}(\theta_{1}-\theta_{2}),\\
\delta\lambda & =\frac{1}{8\,\tau}(\theta_{1}-\theta_{2}).
\end{align}
Here $m$ is a random mass term coming from the real part of the self
energy while $\delta m$ is the analogue term coming from the Imaginary
part of the Self energy. Finally, $\delta\lambda$ is the imaginary
part of the Rashba coupling introduced by the disorder and $\eta$
is the broadening. The quasi-particle lifetime is defined in the main
text as $1/2\tau=n_{i}\,\epsilon\,\alpha_{0}^{2}/4\,v^{2}$. The components
of the disorder averaged propagator now read 
\begin{align}
\mathcal{G}_{\chi,00}^{R/A}(\mathbf{k}) & =-\frac{1}{2}\left\{ \mathcal{L}_{1,\chi}\,[\epsilon+\chi\,(\lambda\pm\imath\,\delta\,\lambda)\pm\imath\,\eta]+\mathcal{L}_{2,\chi}\,[\epsilon-\chi\,(\lambda\,\pm\imath\,\delta\,\lambda)\pm\imath\,\eta]\right\} \\
\mathcal{G}_{\chi,01}^{R/A}(\mathbf{k}) & =-\chi\,\frac{k\,v}{2}\sin(\phi)\left\{ \mathcal{L}_{1,\chi}-\mathcal{L}_{2,\chi}\right\} \\
\mathcal{G}_{\chi,02}^{R/A}(\mathbf{k}) & =\chi\,\frac{k\,v}{2}\cos(\phi)\left\{ \mathcal{L}_{1,\chi}-\mathcal{L}_{2,\chi}\right\} \\
\mathcal{G}_{\chi,10}^{R/A}(\mathbf{k}) & =-\chi\,\frac{k\,v}{2}\cos(\phi)\left\{ \mathcal{L}_{1,\chi}+\mathcal{L}_{2,\chi}\right\} \\
\mathcal{G}_{\chi,20}^{R/A}(\mathbf{k}) & =-\chi\,\frac{k\,v}{2}\sin(\phi)\left\{ \mathcal{L}_{1,\chi}+\mathcal{L}_{2,\chi}\right\} \\
\mathcal{G}_{\chi,33}^{R/A}(\mathbf{k}) & =\frac{\chi}{2}\,(m\,\pm\,\imath\,\delta m)\left\{ \mathcal{L}_{1,\chi}+\mathcal{L}_{2,\chi}\right\} -\frac{\chi}{2}(\lambda\pm\imath\,\delta\lambda)\left\{ \mathcal{L}_{1,\chi}-\mathcal{L}_{2,\chi}\right\} 
\end{align}

\begin{align}
\mathcal{G}_{\chi,11}^{R/A}(\mathbf{k}) & =-\frac{\sin(2\phi)}{4}\{\mathcal{L}_{1,\chi}[\epsilon-\chi\,(m\pm\imath\,\delta m)+2\,\chi\,(\lambda\pm\imath\,\delta\lambda)\pm\imath\,\eta]\\
 & -\mathcal{L}_{2,\chi}[\epsilon-\,\chi\,(m\pm\,\imath\,\delta m)-2\,\chi\,(\lambda\,\pm\,\imath\,\delta\lambda)\pm\,\imath\,\eta]\}\nonumber \\
\mathcal{G}_{\chi,22}^{R/A}(\mathbf{k}) & =-\mathcal{G}_{\chi,11}^{R/A}(\mathbf{k})\\
\mathcal{G}_{\chi,12}^{R/A}(\mathbf{k}) & =\frac{\cos(2\phi)}{4}\{\mathcal{L}_{1,\chi}[\epsilon-\chi\,(m\pm\imath\,\delta m)+2\,\chi\,(\lambda\,\pm\,\imath\,\delta\lambda)\pm\,\imath\,\eta]\\
 & -\mathcal{L}_{2,\chi}[\epsilon-\chi\,(m\,\pm\,\imath\,\delta\,m)-2\,\chi\,(\lambda\pm\,\imath\,\delta\lambda)\pm\,\imath\,\eta]\}+\frac{1}{4}(\epsilon+\chi\,(m\,\pm\imath\,\delta m)\pm\,\imath\,\eta)\{\mathcal{L}_{1,\chi}-\mathcal{L}_{2,\chi}\}\nonumber \\
\mathcal{G}_{\chi,21}^{R/A}(\mathbf{k}) & =\frac{\cos(2\phi)}{4}\{\mathcal{L}_{1,\chi}[\epsilon-\chi\,(m\pm\imath\,\delta m)+2\,\chi\,(\lambda\pm\imath\,\delta\lambda)\pm\,\imath\,\eta]\\
 & -\mathcal{L}_{2,\chi}\,[\epsilon-\,\chi\,(m\pm\,\imath\,\delta m)-2\,\chi\,(\lambda\pm\imath\,\delta\,\lambda)\pm\,\imath\,\eta]\}-\frac{1}{4}(\epsilon+\,\chi\,(m\pm\,\imath\,\delta m)\pm\,\imath\,\eta)\{\mathcal{L}_{1,\chi}-\mathcal{L}_{2,\chi}\}\nonumber 
\end{align}
We have defined the two ''Kernels'' 
\begin{align}
\mathcal{L}_{1,\chi} & =\frac{1}{v^{2}\,k^{2}+(m\pm\imath\,\delta m)[(m\pm\imath\,\delta m)-2(\lambda\pm\imath\,\delta\lambda)]-(\epsilon\pm\imath\,\eta)(\epsilon+2\,\chi\,(\lambda\pm\,\imath\,\delta\lambda)\pm\,\imath\,\eta)}\label{ap:kern}\\
\mathcal{L}_{2,\chi} & =\frac{1}{v^{2}\,k^{2}+(m\pm\imath\,\delta m)[(m\pm\imath\,\delta m)+2(\lambda\pm\imath\,\delta\lambda)]-(\epsilon\pm\imath\,\eta)(\epsilon-2\,\chi\,(\lambda\pm\imath\,\delta\lambda)\pm\imath\,\eta)}
\end{align}
As in the clean case, also in this case a symmetry under valley interchange
exists. Note the form of the disorder-averaged propagators presented
above can be generalised to the non-Gaussian treatment ($T$-matrix)
considering that the matrix structure of the Self-energy is unaffected.
Therefore it suffices to use the corresponding form in the $T$-matrix
approximation of the parameters in Eqs.\,(18)-(21). 

\section{Absence of Schwinger Terms}

In Eq.~(3) of the main text we mentioned the possible existence of
anomalous current commutators in Dirac theories. These commutators
arise from the presence of the infinite Dirac sea and they are at
the core of anomalies. For example, gauge anomalies in (1+1) and (2+1)
are known respectively as chiral and parity anomaly, see Refs.~\cite{bert,zee,frad}
for different aspects and realization of such anomalies. Here we consider
the commutator of the spin density with a charge current. Using the
definition of the spin and current densities: $J_{\mu}^{a}(x)=\psi^{\dagger}(x)\,s^{a}\,\sigma_{\mu}\,v_{\mu}/2\,\psi(x)$
and $J_{\mu}(x)=\psi^{\dagger}(x)\sigma_{\mu}\,v_{\mu}\,\psi(x)$,
where $v_{\mu}=(1,v)$ we can manipulate the commutator as 
\begin{align}
\left[J_{0}^{a}(\mathbf{x}),J_{\mu}(\mathbf{y})\right] & =\frac{1}{2}\left[\psi^{\dagger}(\mathbf{x})\,\gamma_{0}\,s_{a}\,\psi(\mathbf{x}),\psi^{\dagger}(\mathbf{y})\gamma_{\mu}\,\psi(\mathbf{y})\right]\label{eq:HA}\\
 & =\frac{1}{2}\left(\psi^{\dagger}(\mathbf{x})\,\gamma_{0}\,s_{a}\left\{ \psi(\mathbf{x}),\psi^{\dagger}(\mathbf{y})\right\} \gamma_{\mu}\,\psi(\mathbf{y})-\psi^{\dagger}(\mathbf{y})\,\gamma_{\mu}\,\left\{ \psi^{\dagger}(\mathbf{x}),\psi(\mathbf{y})\right\} \gamma_{0}\,s_{a}\,\psi(\mathbf{x})\right)\nonumber \\
 & =\frac{1}{2}\left(\psi^{\dagger}(\mathbf{x})\,s_{a}\,\gamma_{\mu}\,\psi(\mathbf{y})\delta(\mathbf{x-y})-\psi^{\dagger}(\mathbf{y})\,\gamma_{\mu}\,s_{a}\,\psi(\mathbf{x})\,\delta(\mathbf{y-x})\right)\nonumber 
\end{align}
At equal position, the above object is in principle singular as we
are effectively subtracting two infinite quantities~\cite{frad}.
In order to be sure that this term is zero, we need first to regularize
it and then evaluate it explicitly. As a regularization scheme we
choose point splitting with two infinitesimal quantities $\epsilon$
and $\epsilon'$~\cite{frad} and use the normal ordering definition:
$A\,B=\,:A\,B:+\,\langle A\,B\rangle$ %
\begin{align}
 & \left[J_{0}^{a}(\mathbf{x}),J_{\mu}(\mathbf{y})\right]=\lim_{\mathbf{\epsilon},\mathbf{\epsilon}'\to0}\frac{1}{2}\Big(:\psi^{\dagger}(\mathbf{x}+\boldsymbol{\epsilon})\,s_{a}\,\gamma_{\mu}\,\psi(\mathbf{y}-\boldsymbol{\epsilon'}):\delta(\mathbf{x-y}-\boldsymbol{\epsilon-\epsilon'})-:\psi^{\dagger}(\mathbf{y}+\boldsymbol{\epsilon'})\,\gamma_{\mu}\,s_{a}\,\psi(\mathbf{x}-\boldsymbol{\epsilon}):\nonumber \\
 & \times\delta(\mathbf{y-x}-\boldsymbol{\epsilon-\epsilon'})+\langle\psi^{\dagger}(\mathbf{x}+\boldsymbol{\epsilon})\,s_{a}\,\gamma_{\mu}\,\psi(\mathbf{y}-\boldsymbol{\epsilon'})\rangle\,\delta(\mathbf{x-y}-\boldsymbol{\epsilon-\epsilon'})-\langle\psi^{\dagger}(\mathbf{y}+\boldsymbol{\epsilon'})\,\gamma_{\mu}\,s_{a}\,\psi(\mathbf{x}-\boldsymbol{\epsilon})\rangle\nonumber \\
 & \times\,\delta(\mathbf{y-x}-\boldsymbol{\epsilon-\epsilon'})\Big)
\end{align}
where the expectation value is taken with respect to the filled Dirac
sea and $:\,:$ stands for normal ordering. Since the normal ordered
terms are finite, we can now take their difference and so we are left
with the expectation values only. Let us now fix $a=2$ and choose
the gauge such as $E=\partial_{t}\,A_{0}$ \cite{gaugecomm}. It is
also convenient to move to momentum and imaginary frequency space,
from which we arrive at 
\begin{align}
\langle[J_{0}^{2}(q),J_{0}(-p)]\rangle & =-\frac{1}{\beta}\sum_{n}\,\int\frac{d^{2}p}{(2\pi)^{2}}\left\{ G_{02}(\mathbf{p+q},\imath\,\omega_{m}+\imath\,\nu_{n})-G_{02}(\mathbf{p},\imath\,\nu_{n})\right\} ,\label{eq:HA2}
\end{align}
that is the standard form of the Ward Identity. At this point we can
safely shift the momentum in the first Green's function as $\mathbf{p+q}\to\mathbf{p}$
to obtain the cancellation. A longer but equivalent way consists in
performing the integrals explicitly. In this case it is easy to see
that the q-independent Green's function is zero after angular average.
As for the first Green's function, one has to expand for small $q$
and perform the integral explicitly to find again zero. This means
that Eq.~(3) of the main text reduces to a classical conservation
law that completely determines the dynamics of the spin currents,
and in particular the fact that $J_{1/2}^{z}\to0$ as the system reaches
a steady state. 

\section{Details on the evaluation of the spin-Hall conductivity}

\label{sec:SHRen} 

\subsection{Clean system}

Here we show how to obtain the SH conductivity in the clean limit
by a direct evaluation of the SH-response function. The starting point
is the definition of the $\sigma_{SH}$ in terms of its related correlation
function 
\begin{equation}
\sigma_{SH}=\lim_{\omega\to0}\lim_{q\to0}\frac{\Theta_{21}^{3}(\mathbf{q},\omega)}{\imath\,\omega},\label{eq:SF1}
\end{equation}
The Matsubara frequency spin-current response function is 
\begin{align}
\Theta_{21}^{3}(\mathbf{q},\imath\,\omega_{m}) & =-\frac{v^{2}}{2\,\beta}\sum_{n}\int\frac{d^{2}p}{(2\pi)^{2}}\tr\left[(\sigma_{2}\,s_{3})\,G(\mathbf{p+q},\imath\,\nu_{n}+\imath\,\omega_{m})(\sigma_{1}\,s_{0})\,G(\mathbf{p},\imath\,\nu_{n})\right]\label{eq:SF2}\\
 & =-\frac{v^{2}}{2\,\beta}\sum_{n}\int\frac{d^{2}p}{(2\pi)^{2}}\,4\,\imath\,\left\{ G_{33}(\mathbf{p+q},\imath\,\nu_{n}+\imath\,\omega_{m})\,G_{00}(\mathbf{p},\imath\,\nu_{n})-G_{00}(\mathbf{p+q},\imath\,\nu_{n}+\imath\,\omega_{m})\,G_{33}(\mathbf{p},\imath\,\nu_{n})\right\} .\nonumber 
\end{align}
At this point we can take $q\to0$ and evaluate the summation over
Matsubara fermionic frequencies $\nu_{n}$ 
\begin{align}
\Theta_{21}^{3}(0,\imath\,\omega_{m}) & =-\frac{\lambda\,\omega_{m}}{2\,\pi(4\,\lambda^{2}+\omega_{m}^{2})}\int_{|\lambda|}^{v\,\Lambda}dx\left[\frac{2\,x^{2}}{4\,x^{2}+\omega_{m}^{2}}+\frac{2\,\lambda^{2}+\omega_{m}^{2}}{4\,x^{2}+\omega_{m}^{2}}\right]\left\{ f[x-(\epsilon-\lambda)]-f[x-(\epsilon+\lambda)]\right\} ,\label{eq:SF3}
\end{align}
where we have defined $x=\sqrt{v^{2}\,p^{2}+\lambda^{2}}$ and $f[x]$
are fermionic distribution functions. Next we consider the zero temperature
limit of the above expression. There are clearly two different solutions
of the above expression, corresponding to wether $\epsilon$ is greater
or smaller than $\lambda$, i.e. if the Fermi energy intersects two
or one energy bands. It is easy to find in these two regimes 
\begin{align}
\sigma_{SH} & =-\frac{1}{8\pi}\frac{\epsilon^{2}}{\epsilon^{2}-\lambda^{2}}\quad\quad\,\,\,\epsilon>2\lambda\\
\sigma_{SH} & =-\frac{1}{16\pi}\frac{\epsilon(\epsilon+2\,\lambda)}{\lambda(\epsilon+\lambda)}\quad\epsilon<2\lambda.
\end{align}
In order to obtain the above results we have first performed analytic
continuation to real frequencies ($\imath\,\omega_{m}\to\omega+\imath\,0^{+}$)
and then expanded for $\omega<\lambda$. \end{widetext}

}

\begin{thebibliography}{10}
\bibitem{Comment-ExtrinsicSHE}A non-zero SHE in the 2DEG model can
be achieved by extrinsic skew scattering mechanisms. However, this
requires spin\textendash orbit active impurities or breaking time-reversal
symmetry (e.g., by magnetic impurities).
\bibitem{Comment_truncated} By truncated, we mean the vertex which
is irreducible by the cutting of an electron Green function line.
\bibitem{Comment_SG_bubble}Note that $\bar{\sigma}_{\textrm{SG}}$
must not be confused with the full spin density-charge current response
function.
\bibitem{spincomm} Here we use the term spin even though, due to
the Rashba term, spin is not a good quantum number. The two splitted
bands, in each energy sector, are indeed a superposition of spin up
and down.
\bibitem{gaugecomm}It can be checked explicitly that this choice
does not affect the final result.
\end{thebibliography}
\end{document}